\documentclass[letterpaper,10 pt,conference]{ieeeconf}  
\IEEEoverridecommandlockouts                 
\usepackage{mathbbol}
\usepackage{dirtytalk}              
\usepackage{graphicx}
\usepackage{amsmath}
\usepackage{mathtools}
\usepackage{epstopdf}
\usepackage{verbatim}
\usepackage{subcaption}
\usepackage{amssymb}

\newtheorem{definition}{Definition}
\newtheorem{remark}{Remark}
\newtheorem{thm}{Theorem}
\newtheorem{lemma}{Lemma}

\newtheorem{expm}{Example}
\newtheorem{assump}{Assumption}

\title{\LARGE \bf
Towards influence centrality: where to not add an edge in the network?   
}
\author{Aashi Shrinate$^{1}$, \IEEEmembership{Student member, IEEE} and Twinkle Tripathy$^2$, \IEEEmembership{Senior Member, IEEE}
\thanks{$^{1}$Aashi Shrinate is a research scholar and $^{2}$ Twinkle Tripathy is an Assistant Professor in the Control and Automation specialization of the Department of Electrical Engineering, Indian Institute of Technology Kanpur, Kanpur, Uttar Pradesh, India, 208016. Email: {\tt\small aashis21@iitk.ac.in and ttripathy@iitk.ac.in}.}
}

\begin{document}

\maketitle
\thispagestyle{empty}
\pagestyle{empty}

\begin{abstract}
In this work, we consider a strongly connected group of individuals involved in decision-making. The opinions of the individuals evolve using the Friedkin-Johnsen (FJ) model. We consider that there are two competing `influencers' (\textit{stubborn agents}) vying for control over the final opinion of the group. We investigate the impact of modifying the network interactions on their respective control over the final opinions (\textit{influence centrality}). We use signal flow graphs (SFG) to relate the network interactions with the 
influence that each `influencer' exerts on others. We present the sufficient conditions on the edge modifications which lead to the increase of the influence of an `influencer' at the expense of the other. Interestingly, the analysis also reveals the existence of redundant edge modifications that result in no change in the influence centrality of the network. We present several numerical examples to illustrate these results.

\end{abstract}

\section{INTRODUCTION}
\label{Sec:Intro}

An influential node in a social network holds the ability to drive the opinions of others in the networks. In a network with two or more such agents, increasing the influence of one over the other has applications in political campaigns, the promotion of specific brands, combating fake news, \textit{etc.} Such applications rely on suitably varying the influence of a chosen node. Increasing a node's centrality is an intriguing problem that has mainly been explored for PageRank in \cite{avrachenkov2006effect,OLSEN201496} to increase the visibility of a webpage. Recent works have also explored the problem of increasing closeness \cite{improve_closeness} and betweenness centralities \cite{bergamini2018improving} of a node with the objective makes it an efficient spreader of information. 


In social networks, the influence/centrality of a node, along with the network topology, also depends on the underlying process of opinion aggregation \cite{friedkin1991theoretical}. Several models have been proposed in the literature \cite{PROSKURNIKOV201765} that analyse the effect of information aggregation on opinion formation. The FJ model, proposed in \cite{FJ_Model}, is popular due to its simplicity and analytical traceability. In the FJ framework, there are \textit{stubborn agents} who are reluctant to change their opinions and turn out to be influential. Their impact on opinion formation is quantified by the \textit{influence centrality} measure proposed in \cite{Community_Cleavage}. 
In the FJ framework, a random-walk-based interpretation for the final opinions of the agents is proposed in \cite{GHADERI20143209} and  \cite{gionis2013opinion} for undirected and directed graphs, respectively. 

\textit{Relevant literature:} Using the relation between the network topology and the final opinions, network modifications such as edge addition/removal/reweighing of edges are used in the FJ framework to achieve the objectives such as polarization minimisation  \cite{musco2018minimizing,racz2023towards}, reduction of exposure to harmful content \cite{reduce_harm}, opinion maximization \cite{opmax_edge_add}, \textit{etc.}
In general, such network optimisation problems are combinatorial and, therefore, intractable. They employ greedy heuristics due to the non-convex objective functions \cite{racz2023towards,zhu2021minimizing}, NP-hardness of the problem \cite{gionis2013opinion,reduce_harm} \textit{etc}. Recent works \cite{racz2023towards} and \cite{reduce_conflict} use network properties such as Cheeger's constant and the degree of nodes to arrive at closed-form solutions of the impact of an edge addition on the reduction in polarisation and conflict, respectively. Based on this formulation, a disagreement-seeking heuristic is proposed in \cite{racz2023towards} that reduces polarisation by increasing the edge weights of disagreeing nodes. %
The GAMINE algorithm presented in \cite{reduce_harm} 
 reduces exposure to malicious content by utilising the weak connectivity of graphs. It 
substitutes an edge $(i,j)$ with $(i,k)$ to increase connectivity with safe node(s)  that do not have any exposure to harmful content. 
An equivalent approach is implemented in \cite{Sun_Zhang_2023} to achieve opinion maximisation. 
The impact of edge modifications is explored in \cite{controllability_grammian} in the optimisation of controllability grammian-based metrics for an LTI system. A better understanding of the network topology is, therefore, useful in proposing heuristics for network optimisation problems. 

In this work, there are $n$ agents in a strongly connected network whose opinions evolve using the FJ model; two of the agents are stubborn and are competing to maximise their own influence. Our primary objective is to analyse the impact of arbitrary edge modifications on the influence centrality of the network. The proposed analysis has applications in network optimisation problems especially with the objective of opinion maximisation/minimisation \cite{Sun_Zhang_2023} and the reduction of exposure to malicious
content \cite{reduce_harm}. 
In this paper, we consider a modification that mimics the `feed calibration' on social media: an edge $(a,b)$ is added and the edge weight of an existing edge of $(d,b)$ is reduced such that the in-degree of $b$ remains constant \cite{rebalancing_feeds}. Utilising a graphical SFG-based approach, we present topological  conditions on the edge modifications that can change the influence centrality of the network in a desired way, independent of the modified edge weights. Our key contributions are as follows:
\begin{itemize}
    \item We present the sufficient conditions under which an edge modification in the network is redundant. Identifying such edge modifications is relevant to avoid \textit{uneconomical} modifications in practical scenarios. 
    \item Secondly, we present sufficient conditions on the network topology under which an edge modification increases the influence centrality of either of the two stubborn agents \textit{irrespective of edge weights}. 
\end{itemize}

The paper has been organised as follows: Sec. \ref{Sec:Prelims} discusses the relevant preliminaries. We formally define the problem of examining the impact of edge modification 
in Sec. \ref{sec:PS}. Sec. \ref{sec:influence_sfgs} utilises the SFGs to relate an edge modification with the change in influence centrality. Sec. \ref{sec:impact} presents the topology-based conditions that determine the impact of edge modification. Sec. \ref{sec:example} illustrates the results using a suitable example. In Sec. \ref{sec:conclude}, we conclude with insights into the possible future research directions.

\section{Preliminaries}
\label{Sec:Prelims}
\subsection{Graph Preliminaries}
A network of $n$ agents defined as $\mathcal{G}=(\mathcal{V},\mathcal{E})$ where $\mathcal{V}=\{1,2,...,$ $n\}$ is the set of nodes representing the $n$ agents in the network, $\mathcal{E} \subseteq \mathcal{V} \times \mathcal{V}$ is the set of ordered pair of nodes called edges which denote the communication topology of the network. An edge $(i,j)$ denotes the information flow from agent $i$ to agent $j$. The in-neighbours of an agent $i$ is  defined as $\mathcal{N}_{i}^{in}=\{j | (j,i) \in \mathcal{E}\}$ and the out-neighbours of $i$ is defined as  $\mathcal{N}_{i}^{out}=\{j | (i,j) \in \mathcal{E}\}$. A source is a node without in-neighbours and a sink is a node without out-neighbours. 
The matrix 
$W=[w_{ij}] \in \mathbb{R}_{\geq 0}^{n \times n}$ is the weighted adjacency matrix of network $\mathcal{G}$ whose entries $w_{ij}>0$ if an edge $(j,i) \in \mathcal{E}$ else $w_{ij}=0$. The in-degree of a node $i$ is defined as $d_{in}(i)=\sum_{j\in N_{i}^{in}}w_{ij}$.

A path is an ordered sequence of nodes in which every pair of adjacent nodes produces an edge that belongs to the set $\mathcal{E}$.
A \textit{forward path} is one where none of the nodes is traversed more than once. A \textit{feedback loop} is a forward path where the first and the last nodes coincide.
The path gain and loop gain are the product of the edge weights of edges forming the path and loop, respectively. A network is \textit{strongly connected} if each node in the network has a directed path to every other node. 
\subsection{FJ model}
The FJ model is an extension of DeGroot's model that takes stubborn behaviour among agents into account. An agent exhibits stubborn behaviour if it is reluctant to change from its initial position or opinion. 
The opinions of agents in a network $\mathcal{G}$ which are governed by the FJ model evolve as follows,
\begin{align}
\label{eqn:op_model}
\mathbf{x}(k+1)=(I-\beta)W\mathbf{x}(k) + \beta \mathbf{x}(0)  
\end{align}
where $\mathbf{x}(k)=[x_1(k),...,x_n(k)] \in \mathbb{R}^n$ denotes the opinion of $n$ agents in $\mathcal{G}$ at the $k^{th}$ instance, $W$ is a row-stochastic weighted adjacency matrix, $\beta=diag(\beta_1,...,\beta_n)$ is a diagonal matrix with $\beta_i$ denoting the degree of stubbornness of agent $i \in \mathcal{V}$. An agent $i \in \mathcal{V}$ is a stubborn agent if $\beta_i>0$.


\begin{lemma}\cite{FJ_Model}
\label{lemma:FJ_final_op}
  When the underlying network $\mathcal{G}=(\mathcal{V},\mathcal{E})$ is strongly connected and $\beta_i \in [0,1)$ such that $\beta_i>0$ for at least one $i \in \mathcal{V}$, the following conditions hold:
  \begin{itemize}
      \item at steady state,  the opinions of the agents converge to,
\begin{align}
\label{eqn:fin_opinion}
    \mathbf{x}_f=P \mathbf{x}(0)  
\end{align}
where $\mathbf{x}_f=[x_{1f},...,x_{nf}]=\lim_{k \to \infty}\mathbf{x}(k)$ and $P=(I-(I-\beta)W)^{-1}\beta$.
\item the matrix $P$ is row-stochastic
  \end{itemize}
\end{lemma} 

\subsection{Signal Flow Graphs and Index Residue Reduction}
\label{Sec:Index_residue}
An SFG $\mathcal{G}_s=(V_s,E_s)$ is a graph with each node $i\in V_s$ is associated with state $y_i$. Each state $y_i$ satisfies a linear equation of the form $y_i=\sum_{j=1}^{r}g_{i,r}y_r$; where $g_{i,r}$ is the branch gain of branch $(r,i)$ denoting the dependence of state $y_i$ on state $y_r$. 
In an SFG, one or more sources may exist. Note that the state associated with a source is independent of any other state while it directly or indirectly affects the states associated with the other nodes in the SFG.

In general, the SFGs are used to determine the effect of inputs on various states in the system. Each input forms a source in the SFG. The gain of the SFG  is defined as the signal
appearing at the sink per unit signal applied at the
source \cite{mason1956feedback}. 
In the absence of loops, the gain of SFG $\mathcal{G}_s$ for a pair of source and sink is the summation of path gains of all forward paths from the source to the sink in $\mathcal{G}_s$. On the other hand, if an SFG has a self-loop with gain $a$ it is replaced by a branch of gain $1/(1-a)$ and the gain of the SFG is calculated. However, in the presence of loops consisting of two or more nodes, additional terminologies are required which are defined as follows.

\begin{figure}[ht]
    \centering
\begin{subfigure}{0.2\textwidth}
\centering
\includegraphics[width=0.6\textwidth]{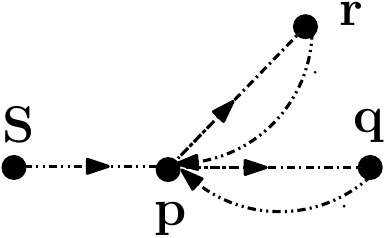}
    \caption{An SFG $\mathcal{G}_s$}
    \label{fig:1_index_node} 
\end{subfigure}%
\begin{subfigure}{0.28\textwidth}
    \centering
    \includegraphics[width=0.5\textwidth]{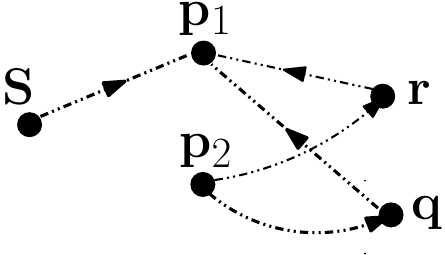}
    \caption{The SFG $\mathcal{G}_s$ after node $p$ is split }
    \label{fig:2_index_node} 
\end{subfigure}
\caption{The SFG in Fig. \ref{fig:1_index_node} has nodes $p,q,r$ and source $S$. Splitting node $r$ into two nodes $r_1$ and $r_2$ 
 eliminates all the loops in $\mathcal{G}_s$ passing through the node as observed in Fig. \ref{fig:2_index_node}.}
 \label{fig:node_split}
\end{figure}

In an SFG, a node $r$ can be split into two nodes $r_1$ and $r_2$, such that node $r_1$ is connected with all the outgoing branches of $r$ and $r_2$ is connected to all incoming branches of $r$. The smallest set of nodes that must be split to remove all the loops in the graph is denoted by $\mathcal{I}$. The nodes in $\mathcal{I}$ 
are referred to as index nodes. In Fig. \ref{fig:node_split}, we split a node $r$ in $\mathcal{G}_s$ given in Fig. \ref{fig:1_index_node}. This results in an SFG without any loops shown in Fig. \ref{fig:2_index_node}. Therefore, it follows that $r$ is an index node.  Note that the index nodes of a graph need not be unique.

 The gain of an SFG $\mathcal{G}_s$ with loops is determined by using index-residue reduction of the SFG. A reduced SFG $\mathcal{G}_s^1$ is formed that has nodes consisting of sources, sinks and index nodes of $\mathcal{G}_s$. The nodes retained in $\mathcal{G}_s^1$ are called residual nodes, and the residual path is a path in $\mathcal{G}_s$ that starts from a residual node to another residual node or (itself) without passing through any other residual node. A branch $(i,j)$ exists in $\mathcal{G}_s^1$ if a residual path exists in $\mathcal{G}_s$ from an index node or a source $i$ to an index node or a sink $j$. The non-residual nodes are the nodes in $\mathcal{G}_s$ other than the residual nodes. 
  The branch gain $g_{i,j}^1$ is equal to the sum of path gains of the residual paths $j \to i$ in $\mathcal{G}_s$. Next, the self-loops, if any, are converted into branches by the aforementioned procedure. If the reduced graph $\mathcal{G}_s^1$ does not contain any loops, we determine the gain of the sources. Otherwise, the index-residue reduction is performed iteratively until the reduced graph does not contain any loops.

\section{Problem Statement}
\label{sec:PS}
The FJ model is one of the most well-studied models in the literature of mathematical sociology and opinion dynamics. This model introduced stubborn behaviour in agents, which results in disagreement, the commonly occurring outcome of social interactions. 
In a network of agents, the term \textit{influential} simply refers to an agent having a say in the final opinions of the agents in the network. Interestingly, when the opinions of the agents evolve by the FJ model, only the stubborn agents are influential \cite{FJ_Model}. Formally, we define an influential agent in the FJ framework as follows:

\begin{definition} A stubborn agent $i \in \mathcal{V}$ is said to influence another agent $j \in \mathcal{V}$, when its initial opinion $x_i(0)$ contributes to the final opinion $\bar{x}_j$ of agent $j$. 
\end{definition}
Once the influential agents are identified, a natural question arises about their impact on the other agents in the network. 
The influence centrality measure, proposed in \cite{Community_Cleavage}, quantifies the contribution of stubborn agent(s) on the final opinions of the agents in the framework of FJ model.
Mathematically, it is defined as follows:
\begin{align}
\label{eqn:influence_centrality}
    \mathbf{c}=\frac{P^T \mathbb{1}_n}{n}
\end{align}
where $\mathbf{c}=[c_1,...,c_n]$ denotes the influence centrality measure of $n$ agents in the network whose entry $c_i$ gives
the degree of influence of a stubborn agent $i$. 
The influence centrality measure satisfies $\mathbf{c}^T\mathbb{1}_n=1$, implying that if the influence of a stubborn agent increases, there exists another stubborn agent whose influence decreases.

It is well known that the influence centrality measure depends on the degree of stubbornness of all the agents and the network topology \cite{FJ_Model,GHADERI20143209}. In practical applications such as the mitigation of the spread of fake news or promoting certain brands to wider target audiences, it might be desirable to make one agent (or brand) more influential (or popular) than the others. Therefore, harping on the critical role of network topology on the degree of influence, \textit{we analyse the effect of edge modifications to suitably alter a stubborn agent's influence in a strongly connected network}. 

\subsection{Implications of edge modifications in social networks} 

\begin{figure}[h]
       \centering
       \begin{subfigure}{0.19\textwidth}
       \centering
\includegraphics[width=0.65\linewidth,keepaspectratio]{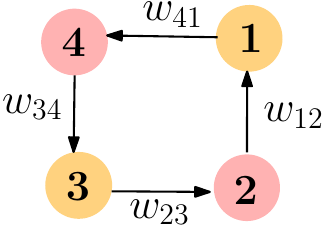}
       \caption{Network $\mathcal{G}$}
       \label{fig:example_network}  
       \end{subfigure}
            \begin{subfigure}{0.19\textwidth}
       \centering
 \includegraphics[width=0.65\linewidth,height=2.2cm,keepaspectratio]{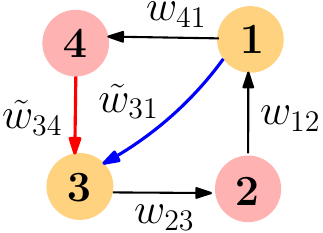}
       \caption{Modified network $\mathcal{\hat{G}}$ }
       \label{fig:example_network_Edge_Add}  
       \end{subfigure}   
       \caption{Effect of the \textit{constant in-degree} condition.}
   \end{figure}

The addition of a directional edge $(a,b)$ between two nodes $a$ and $b$ in a social network can be interpreted as a suggestion for $b$ to establish a friendship with $a$ and follow its posts. Adding new friends on social media leads to a phenomenon of `feed alteration' where the visibility of posts from existing friends may decrease to accommodate the content from the newly added ones. Therefore, \textit{the added edge $(a,b)$ is accommodated by a reduction in edge weight of an existing edge, say $(d,b)$ such that the in-degree of $b$ remains constant.} Formally, we define an edge modification as:
\begin{definition}
    Given distinct nodes $a$, $b$ and $d$, an edge modification $(a,b,d)$ is the addition of an edge $(a,b)$ and the reduction of edge weight of the existing edge $(b,d)$ such that the \textit{in-degree of $b$ remains constant} which implies:
\begin{align}
\label{eqn:weight_condition}
    w_{bd}=\tilde{w}_{ba}+\tilde{w}_{bd}
\end{align}
where $w_{bd}$ is the weight of the edge $(d,b)$ in $\mathcal{G}$ and $\tilde{w}_{ba}$, $\tilde{w}_{bd}$ are the weights of the edges $(a,b)$ and $(d,b)$, respectively, after the edge modification such that $\tilde{w}_{ba}>0$ and $\tilde{w}_{bd}>0$.
\end{definition}
%
Note that, throughout the paper, the added edge is indicated by a blue arrow and the edge whose weight is reduced is represented by a red arrow.
\begin{expm}
 Consider the network of $4$ agents shown in Fig. \ref{fig:example_network}. The set of stubborn agents, $\{2,4\}$, are denoted by the nodes in red colour. An edge modification $(1,3,4)$ is applied on the network $\mathcal{G}$ as shown in Fig. \ref{fig:example_network_Edge_Add}. To maintain a \textit{constant in-degree} for node $3$, we must have $w_{34}=\tilde{w}_{31}+\tilde{w}_{34}$. The rest of the edge weights remain unchanged.  
\end{expm}

\subsection{Primary objectives}
In this paper, our primary objective is to analyse the impact of edge modifications on the degree of influence of the stubborn agents. 
For simplicity, we consider only two agents in the network to be stubborn. Therefore, an increase in the influence of either of the stubborn agents results in a decrease in the influence of the other. 
Such a framework captures real-world contests for influence like in a bipartisan political system or a duopoly of firms. Mathematically, we are interested in arriving at sufficient conditions on the network topology such that the edge modification $(a,b,d)$ results in a suitable change in the degree of influence of \textit{only} either one of the stubborn agents.
\section{Influence and signal flow graphs}
\label{sec:influence_sfgs}
Consider a network $\mathcal{G}=(\mathcal{V},\mathcal{E})$ of agents whose opinions evolve by eqn. \eqref{eqn:op_model} when only two agents in the network are stubborn. In the given framework, the final opinions of the agents in the network satisfy the following set of linear equations,
\begin{align}
\label{eqn:sfg}
    \mathbf{x}_f=(I-\beta)W\mathbf{x}_f+\beta\mathbf{{x}}(0)
\end{align}
The average of final opinions of agents, denoted by $\bar{x}$, relates to the influence centrality measure by the following equations.
\begin{align}
\label{eqn-average_op}
\bar{x}=\frac{{\mathbf{x}_f}^T\mathbb{1}_n}{n}= \frac{\mathbf{x}(0)^T P^T \mathbb{1}_n}{n}= \mathbf{x}(0)^T \mathbf{c}
\end{align}
In the given framework, 
$\bar{x}$ depends only on the initial opinions of stubborn agents. It follows from eqn. \eqref{eqn-average_op} that the contribution of a stubborn agent $i$ in $\bar{x}$ is equal to its degree of influence $c_i$.
%
We concatenate eqns. \eqref{eqn:sfg} and \eqref{eqn-average_op} to get the following relation. 
%
\begin{align}
\label{eqn:sfg_2}
   \begin{bmatrix}
       \mathbf{x}_f \\
       \bar{x}
   \end{bmatrix}= \begin{bmatrix}
    (I-\beta)W\mathbf{x}_f+\beta\mathbf{{x}}(0) \\
    \frac{\mathbb{1}_n^T \mathbf{x}_f}{n}
    \end{bmatrix}
\end{align}
This steady-state relation can be expressed graphically using an SFG. \textit{Such a representation is useful in visualising how the final opinions of different agents are related to one another.} (See \cite{shrinate2024absolute} for a detailed description of the use of SFG in influence analysis.)

In the given framework,
the SFG $\mathcal{G}_s=(V_s,E_s)$, derived using \eqref{eqn:sfg_2}, is composed of the following nodes:
\begin{itemize}
    \item Each of the first $n$ nodes represent the final opinion $x_{if}$ of the $i^{th}$ agent in the network. 
    \item The next two nodes correspond to the initial opinion $x_i(0)$ of the stubborn agent $i$ in $\mathcal{G}$. These nodes form sources $S_1$ and $S_2$ in $\mathcal{G}_s$.
    \item The last node is associated with average final opinion $\bar{x}$. It forms a sink in $\mathcal{G}_s$ and is denoted by $O$.
\end{itemize}
where $i\in\{1,2,..,n\}$.
We use the relations in eqn. \eqref{eqn:sfg_2} to form the branches in $\mathcal{G}_s$.
A branch $(i,j)$ in $\mathcal{G}_s$ has branch gain $g_{i,j}=(1-\beta_i)w_{ij}$ for $i,j \in \{1,...,n\}$. Further, 
sources $S_1$ and $S_2$ in $\mathcal{G}_s$ have out-going branches with branch gains being their degrees of stubbornness. The 
sink $O$ in $\mathcal{G}_s$ has an incoming branch $(i,O)$ from each node $i$ in $\mathcal{G}_s$ with branch gain $1/n$ where $i \in \{1,...,n\}$. Using the given procedure, the SFG constructed for a strongly connected network in Fig. \ref{fig:example_network} is shown in Fig. \ref{fig:example_sfg}.

The degree of influence $c_i$ of a stubborn agent $i$ is the gain of the SFG for source $S_j$ and sink $O$ where $S_j$ is associated with $x_i(0)$,  $i \in \{1,...,n\}$ and $j \in \{1,2\}$. In this paper, our aim is to determine the change in the degree of influence $c_i$ resulting from an edge modification in $\mathcal{G}$ and attribute its underlying causes. Therefore, we use the 
index residue reduction (detailed in Appendix A) to further simplify the SFG. 

\subsection{Impact of edge modification in  $\mathcal{G}$ on $\mathcal{G}_s$ }
It is simple to observe that the addition of an edge $(a,b)$ in the network $\mathcal{G}$ is equivalent to the addition of a branch $(a,b)$ in the SFG $\mathcal{G}_s$ such that $a,b \in \{1,...,n\}$. Similarly, the reduction of edge weight of $(d,b)$ in $\mathcal{G}$ results in the reduction of branch gain of $(d,b)$ in $\mathcal{G}_s$ where $d \in \{1,...,n\}$. It can be determined that the in-degree of any node in $\mathcal{G}_s$ except sources is equal to $1$, which further remains constant despite the edge modification. 
The impact of edge modification in $\mathcal{G}$ on $\mathcal{G}_s$ is demonstrated in Example \ref{expm:2}. 
%
%
\begin{expm}
\label{expm:2}
For the network $\mathcal{G}$ in Fig. \ref{fig:example_network}, we construct its SFG as shown in Fig. \ref{fig:example_sfg}. The network in Fig. \ref{fig:example_network_Edge_Add} is obtained after an edge modification $(1,3,4)$.
Equivalent changes are reflected in the SFG $\mathcal{G}_s$ as shown in Fig. \ref{fig:example_sfg_edge_Add}. 

   \begin{figure}[h]
       \centering
       \begin{subfigure}{0.23\textwidth}
       \centering
         \includegraphics[width=1\linewidth,height=2.0cm,keepaspectratio]{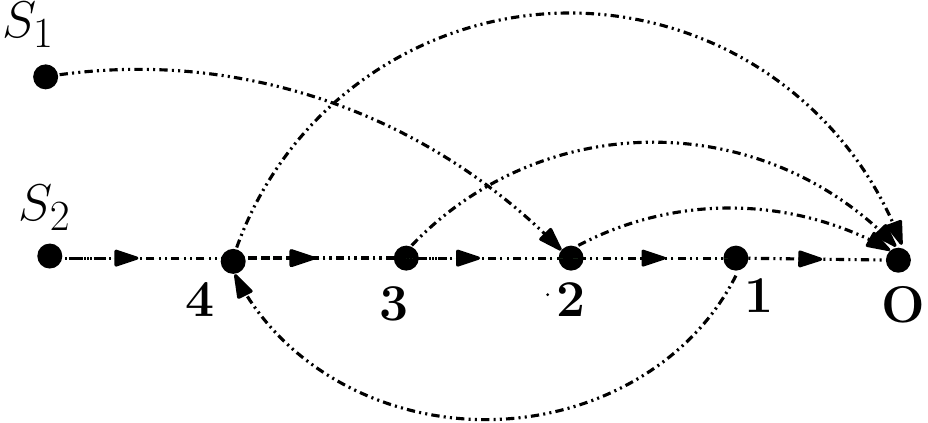}
       \caption{SFG $\mathcal{G}_s$}
       \label{fig:example_sfg}  
       \end{subfigure}
           \begin{subfigure}{0.23\textwidth}
       \centering
    \includegraphics[width=1\linewidth,height=2.0cm,keepaspectratio]{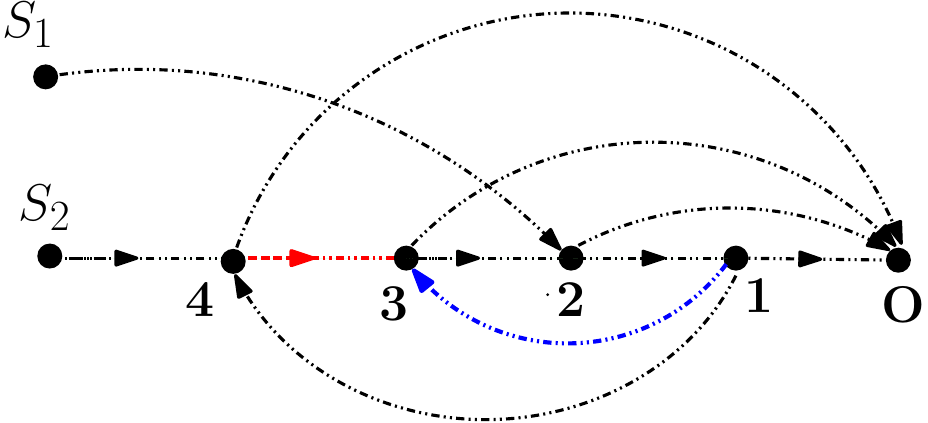}
       \caption{Modified SFG $\mathcal{G}_s$}
       \label{fig:example_sfg_edge_Add}  
       \end{subfigure}
       \caption{Effect of the edge modification $(1,3,4)$ in $\mathcal{G}$ on its corresponding SFG $\mathcal{G}_s$}
       \label{fig:enter-label}
   \end{figure}
\end{expm}
\subsection{Index nodes and the permissible edge modifications}
The analysis of the impact of edge modifications is extremely useful in real-world applications, an example being the marketing of products in appropriate or selective regions within a network. However, the task becomes increasingly challenging as the networks become more dense with edges. Keeping this in mind, in this paper, we consider a subset of strongly connected networks that adhere to the following assumption.
\begin{assump}
\label{Assump:index}
Given a network $\mathcal{G}$, there exists a node that occurs in every loop in the $\mathcal{G}$ except the self-loops.
\end{assump}

Due to Assumption \ref{Assump:index}, an interesting property arises for the underlying SFG $\mathcal{G}_s$ corresponding to the network $\mathcal{G}$. In general, any SFG can be subjected to a procedure known as the \textit{index residue reduction}. This procedure starts with the identification of the \textit{index nodes} in the network. Loosely speaking, an index node is one which is associated with one or more loops in the network; \textit{interestingly, the network loses its cyclicity in the absence of all of its index nodes}. The complete procedure of the identification of index nodes and the index residue reduction is detailed in Sec. \ref{Sec:Index_residue}.
\begin{remark}
\label{remark_one_index_node}
    For a given network $\mathcal{G}$ that satisfies Assumption \ref{Assump:index}, the corresponding SFG $\mathcal{G}_s$ has only one index node. 
\end{remark} 

\begin{figure}[h]
    \centering
    \includegraphics[width=0.8\linewidth,keepaspectratio]{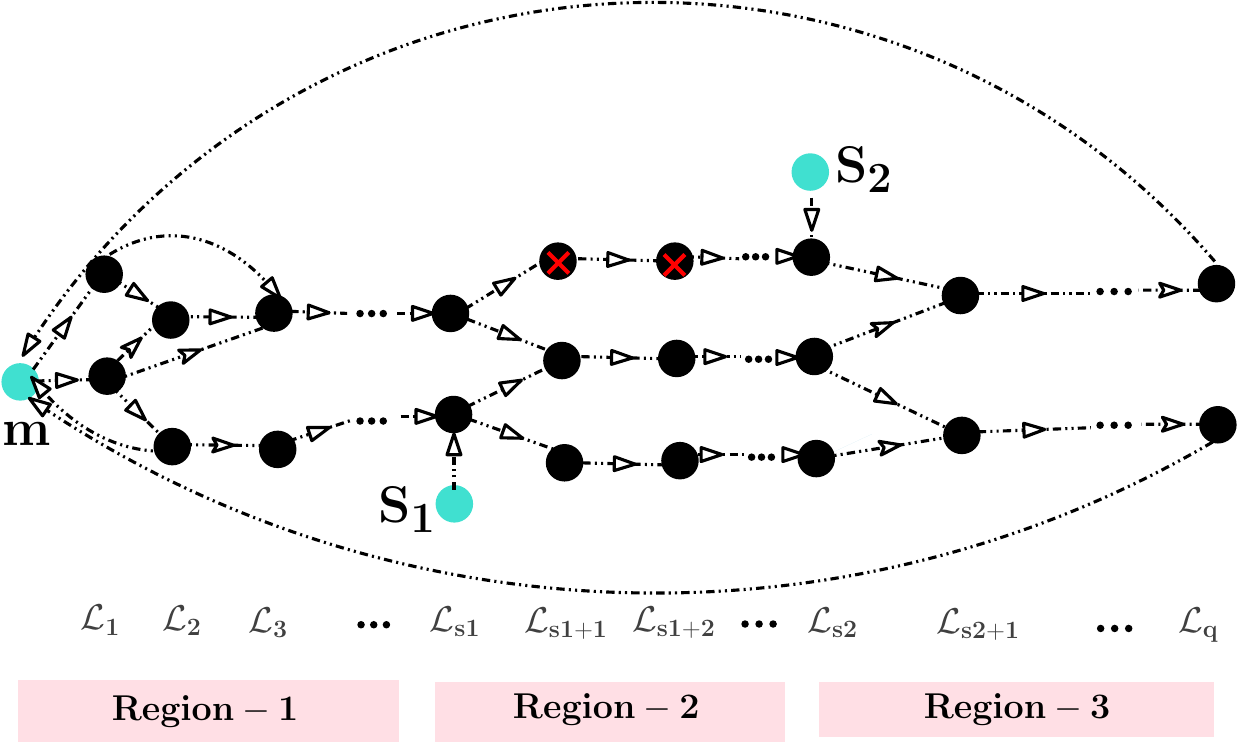}
    \caption{{The distribution of non-residual nodes in level sets.}}
    \label{fig:Lvel_distribution}
\end{figure}

\begin{expm}
\label{expm:3}
The network $\mathcal{G}$ shown in Fig. \ref{fig:example_network_Edge_Add} has two loops  $1\to 2 \to 3 \to 4 \to 1$ and $1 \to 3 \to 2 \to 1$. Nodes $1,2$ and $3$ are common in both the loops such that if we split one of the corresponding nodes in the SFG $\mathcal{G}_s$, then $\mathcal{G}_s$ becomes acyclic. Thus, either one of $ 1,  2$ or  $3$ can be chosen as an index node. 
\end{expm}
\begin{lemma}
\label{lemma:disjoint_levels}
In a network $\mathcal{G}$ satisfying Assumption \ref{Assump:index}, the non-residual nodes (the nodes except the index node, sink and sources) in $\mathcal{G}_s$ 
can be distributed into disjoint sets of levels $\mathcal{L}_1,...,\mathcal{L}_q$ where $q\leq n-1$. A node $j$ in a set $\mathcal{L}_h$ satisfies the following conditions on its in-neighbours:
\begin{enumerate}
    \item the nodes in the set $\mathcal{L}_1,...,\mathcal{L}_{h-1}$ are allowed to be its in-neighbours, and,
    \item it must have at least one in-neighbour from $\mathcal{L}_{h-1}$.
   \end{enumerate}
\end{lemma}
\begin{proof}
During the construction of the SFG $\mathcal{G}_s$, WLOG, we choose to make the index node, say $m$, as the first node. The reason is its connectivity with the other nodes. Due to the Assumption \ref{Assump:index}, the non-residual nodes cannot form a loop that does not have node $m$.  
If a node in $\mathcal{L}_h$ has in-neighbours from a node in set $\mathcal{L}_{h+1},...,\mathcal{L}_{q}$, then the non-residual nodes may form a loop amongst themselves. This violates Assumption \ref{Assump:index} leading to a contradiction. Thus, condition (1) holds. Condition (2) holds so that each non-residual node belongs to a unique set $\mathcal{L}_h$. 
\end{proof}
The non-residual nodes in $\mathcal{G}_s$ can be distributed in levels as shown in Fig. \ref{fig:Lvel_distribution}. In $\mathcal{G}_s$, the nodes highlighted in blue in Fig. \ref{fig:Lvel_distribution}, denote the residual nodes.
For ease of visualisation, a source in $\mathcal{G}_s$ associated with the initial opinion of a stubborn agent is placed in the same level set as the node associated with the final opinion of the stubborn agent. The source whose appearance precedes the other one in the levels is termed as $S_1$ and the other one is referred to as $S_2$. The sets $\mathcal{L}_{s1}$, $\mathcal{L}_{s2}$ are the ones which contain a node with an incoming edge from $S_1$ and $S_2$, respectively. Note that the sink $O$ is omitted in Fig. \ref{fig:Lvel_distribution} to avoid confusion. 

\begin{figure}[h]
    \centering
    \begin{subfigure}{0.23\textwidth}
       \centering
\includegraphics[width=0.9\linewidth,height=2.5cm,keepaspectratio]{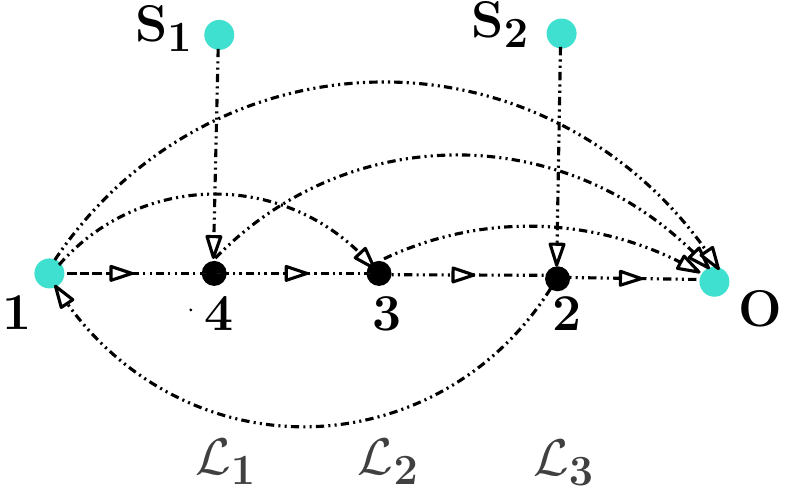}
       \caption{Node $1$ as index node}
       \label{fig:Level_1}  
       \end{subfigure}
       \begin{subfigure}{0.23\textwidth}
       \centering
\includegraphics[width=0.9\linewidth,height=2.5cm,keepaspectratio]{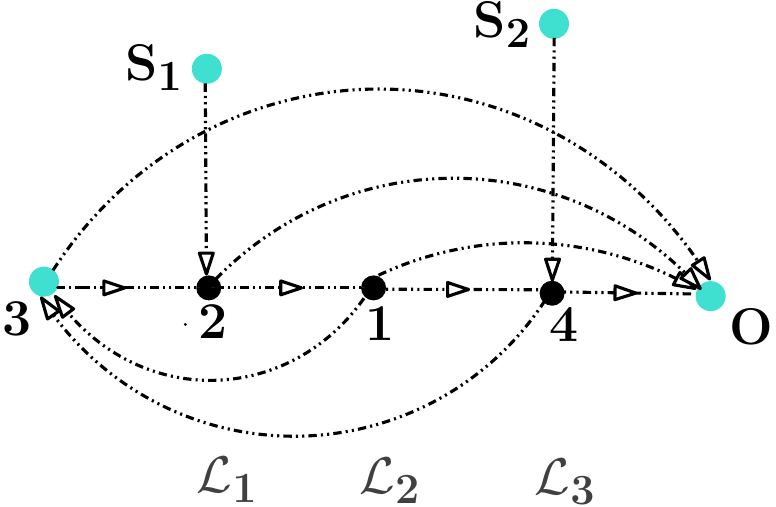}
       \caption{Node $3$ as index node}
       \label{fig:Level_2}  
       \end{subfigure}
    \caption{The role of index nodes in the level sets' construction.}
    \label{fig:enter-label}
\end{figure}

An important point to note is that the distribution of nodes into levels sets depends on the choice of the index node as explained next. 
\begin{expm}
\label{expm:4}
   We distribute the non-residual nodes of the SFG in Fig. \ref{fig:example_sfg_edge_Add} into level sets. We know from Example \ref{expm:3} that the SFG has nodes $1,2$ and $3$ as eligible index nodes. We distribute the non-residual nodes of the SFG by considering node $1$ as the index node in Fig. \ref{fig:Level_1} and 
   by considering node $3$ as the index node in Fig. \ref{fig:Level_2}.
   Comparing the Figs. \ref{fig:Level_1} and \ref{fig:Level_2}, we observe that the members of level sets differ with the choice of the index node. Furthermore, note that in Fig. \ref{fig:Level_1}, 
   $S_1$ corresponds to the initial opinion of $4$ and $S_2$ corresponds to the initial opinion of $2$. On the other hand, in Fig. \ref{fig:Level_2}, the vice-versa occurs. 
\end{expm}

Example \ref{expm:4} illustrates the role of the index node in the distribution of nodes into level sets. 
Next, we present the conditions for permissible edge modifications in this framework.
Throughout this work, we consider the edge modification applied to a network $\mathcal{G}$ satisfies the following assumption.
\begin{assump}
\label{assump:network_modifications}
 Given a network $\mathcal{G}$, an edge modification $(a,b,d)$ is permissible if the modified network satisfies Assumption \ref{Assump:index}.  
\end{assump}

Assumption \ref{assump:network_modifications} implies that an edge modification is permissible only when the modified network has atmost one index node. We can visualise the conditions for a permissible edge modification $(a,b,d)$ in terms of the level sets defined for a given index node $m$. Due to the edge addition of $(a,b)$ where $a \in \mathcal{L}_w$ and $b \in \mathcal{L}_v$, two cases might arise (i) $w\leq v$ and (ii) $w>v$.
In the first case, the edge addition is always permissible because conditions in Lemma \ref{lemma:disjoint_levels} hold and $m$ continues to be the index node. On the other hand, when $w>v$, it may result in the formation of a loop without the index node $m$, implying $m$ is no longer the index node participating in every loop in the network. Therefore, such edge addition is permissible only when another eligible index node $m'$ exists such that the Assumption \ref{Assump:index} holds.
Having explored the impact of the edge modifications on $\mathcal{G}_s$, we analyse their impact on the influence centrality of a network. We show that the use of the SFGs largely simplifies the analysis in the given framework.

\section{Impact of edge Modifications on Influence}
\label{sec:impact}
As discussed before, in this paper, we consider a strongly connected network $\mathcal{G}$ consisting of two stubborn agents. Our primary objective is to analyse the change in the influence of the stubborn agents due to permissible edge modifications. A natural question that arises is: \textit{which edge modifications in the network would increase the influence of a selected stubborn agent over the other?} In this section, we try to answer this question through the following results.

\begin{definition}
     A node $i$ is said to have a \textit{direct path} from a node $j$, if a path $j \to i$ exists in $\mathcal{G}_s$ which does not pass through the index node $m$ for $i,j,m \in V_s$ such that both $i$ and $j$ are not residual nodes.
\end{definition}
If both $i$ and $j$ are residual nodes in $\mathcal{G}_s$, then a path $i \to j$ that does not pass through $m$ is a residual path and not a direct path. 
\begin{thm}
\label{cor_1}
Consider a strongly connected network $\mathcal{G}$ where an edge modification $(a,b,d)$ is applied such that Assumptions \ref{Assump:index} and \ref{assump:network_modifications} hold. In $\mathcal{G}_s$, if an index node $m$ exists such that neither $a$ nor $d$ have a direct path from the sources $S_1$ or $S_2$ (defined with respect to $m$), then 
the degree of influence of the stubborn agents does not change. 
\end{thm}

The proof of Theorem \ref{cor_1} is presented in Appendix B. We use the level sets to visualise the conditions given in Theorem \ref{cor_1}. 
As shown in Fig. \ref{fig:Lvel_distribution}, the nodes in $\mathcal{G}_s$ can be distributed in three regions: (1) the nodes from the sets $\mathcal{L}_1$ to the
 set $\mathcal{L}_{s1-1}$ including $m$, (2) the nodes in sets from $\mathcal{L}_{s1}$ to the set $\mathcal{L}_{s2-1}$ and (3) the nodes in sets from $\mathcal{L}_{s2}$ to $\mathcal{L}_q$. 

For a permissible edge modification $(a,b,d)$ such that $a$ and $d$ lie in region $1$, it follows from Lemma \ref{lemma:disjoint_levels} that every path from the sources $S_1$ and $S_2$ passes through $m$. Thus, the condition in Theorem \ref{cor_1} holds, and such edge modifications are redundant.
Even in regions $2$ and $3$, there may exist certain nodes that
do not have a path from $S_1$ without passing through $m$. For example, observe the nodes in Fig. \ref{fig:Lvel_distribution} in the region $2$ marked with red crosses. For such $a$ and $d$ also, the $(a,b,d)$ is also redundant. Since no constraints are imposed on the position of $b$, it can lie in any region as long as  $(a,b)$ is a permissible edge addition.
It follows from Theorem \ref{cor_1} that edge modifications such that both $a$ and $d$ lie in region $1$ are redundant. Next, we consider the edge modifications where either one of $a$ or $d$ lies in the region $2$ and has a direct path from $S_1$. 

\begin{thm}
\label{Cor_two}
Consider a strongly connected network $\mathcal{G}$ where an edge modification $(a,b,d)$ is applied such that 
the Assumptions \ref{Assump:index} and \ref{assump:network_modifications} hold. Additionally, let the following conditions hold for an index node $m$ in $\mathcal{G}_s$:
\begin{itemize}
    \item $a$ does not have a direct path from any of the sources. 
    \item $d$ has at least one direct path from a source ${S}_1$ 
\end{itemize}
The influence of the stubborn agent corresponding to $S_1$ decreases under these conditions. If the nodes $a$ and $d$ are interchanged, then the influence of the stubborn agent corresponding to $S_1$ increases. 
\end{thm}
The proof for the Theorem \ref{Cor_two} is presented in Appendix B.
\begin{remark}
   Intuitively, an edge modification $(a,b,d)$ such that $a$ and $d$ have no direct paths from $S_1$ and $S_2$ is redundant because the effect of change gets distributed among both the sources. 
    On the other hand, when $a$ 
    has a direct path from $S_1$, and $d$ does not have a direct path from any source. Then, the newly added edge $(a,b)$ lies on forward path(s) only from $S_1$, while the reduced edge $(d,b)$ lies on paths from both $S_1$ and $S_2$. Therefore, it follows that the influence of $S_2$ decreases leading to an increase in the influence of $S_1$.
    
\end{remark}
\section{An Illustrative example}
\label{sec:example}
\begin{figure}[h]
       \centering
       \begin{subfigure}{0.11\textwidth}
       \centering
\includegraphics[width=0.9\linewidth,height=2.2cm,keepaspectratio]{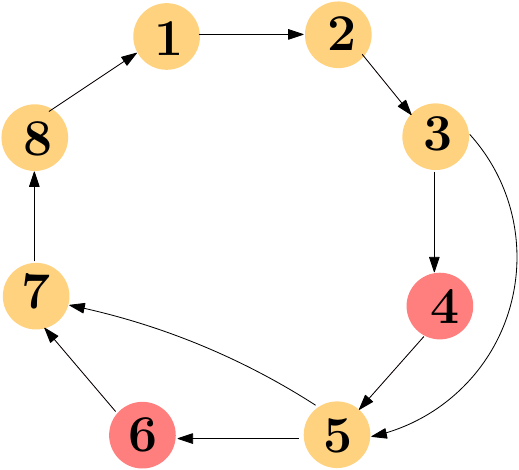}
       \caption{Network $\mathcal{G}$}
       \label{fig:Case_1}  
       \end{subfigure}
       \begin{subfigure}{0.11\textwidth}
       \centering
\includegraphics[width=0.9\linewidth,height=2.2cm,keepaspectratio]{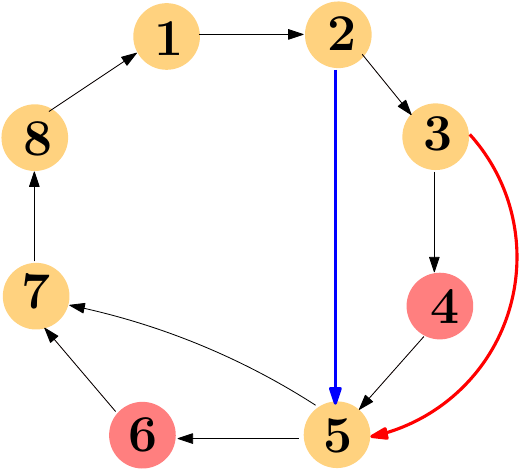}
       \caption{$(2,5,3)$}
       \label{fig:Case_1_1}  
       \end{subfigure}
       \begin{subfigure}{0.11\textwidth}
       \centering
\includegraphics[width=0.9\linewidth,height=2.2cm,keepaspectratio]{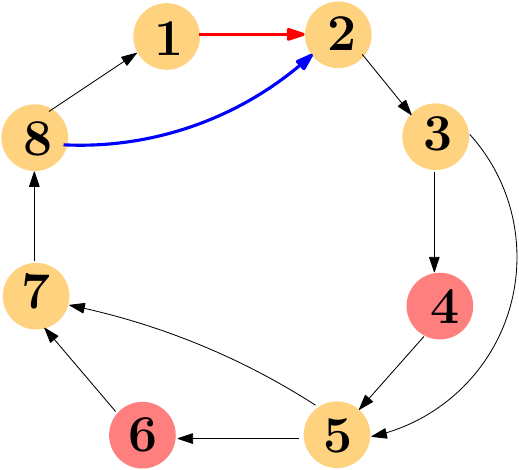}
       \caption{$(8,2,1)$}
       \label{fig:Case_1_2}  
       \end{subfigure}
       \begin{subfigure}{0.11\textwidth}
       \centering
\includegraphics[width=0.9\linewidth,height=2.2cm,keepaspectratio]{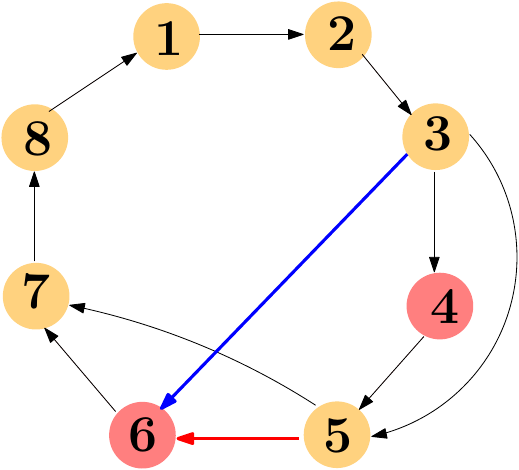}
       \caption{$(3,6,5)$}
       \label{fig:Case_1_2}  
       \end{subfigure}
        \caption{Edge modifications in $\mathcal{G}$}
       \label{fig:example_1}
\end{figure}

Consider the network $\mathcal{G}$ shown in Fig. \ref{fig:Case_1} where agents $4$ and $6$ are stubborn. We assume $\mathcal{G}$ has no self-loops and the weight of each edge is equal. The degree of stubbornness of $4$ and $6$ is $0.1$ and $0.3$, respectively. The degree of influence is determined using eqn. \eqref{eqn:influence_centrality}, turns out to be $c_4=0.30$ and $c_6= 0.70$. The nodes $1,2,3,5,7$ and $8$ in $\mathcal{G}_s$ are eligible to be the index nodes. We choose $1$ as the index node and introduce an edge modification $(2,5,3)$ shown in Fig. \ref{fig:Case_1_1}. The weight of the added edge is $w_{52}=0.3$.  In $\mathcal{G}_s$, nodes $2$ and $3$ do not have direct paths from either of the sources (because in $\mathcal{G}$ every path from stubborn agents $4$ or $6$ to agents $2$ or $3$ passes through $1$). The degree of influence in the modified network is unchanged and turns out to be $c_4=0.30$ and $c_6=0.70$. 

Another permissible edge modification $(8,2,1)$ is applied to $\mathcal{G}$, as shown in Fig. \ref{fig:Case_1_2}. The weight $w_{28}=0.5$. Note that node $1$ is no longer an index node. However, for the choice of node $7$ as the index node, the Assumption \ref{assump:network_modifications} holds, and the nodes $8$, $1$  do not have direct paths from sources. The degree of influence of the modified network verifies that the edge modification is redundant as $c_4=0.3$ and $c_6=0.7$. 
Next, we make an edge modification $(3,6,5)$ with $w_{63}=0.5$ in the network $\mathcal{G}$. Consider $1$ as index node, the node $3$ does not have a direct path from sources but $5$ has a direct path from $S_1$ associated with stubborn agent $4$. Therefore, according to Theorem \ref{Cor_two}, the influence of stubborn agent $4$ decreases and that of $6$ increases. The degree of influence of the modified network is $c_4=0.26$ and $c_6=0.74$. Hence, verified.

\section{CONCLUSIONS}
\label{sec:conclude}
In this paper, we examine the impact of edge modifications on the influence of the stubborn agents in a network governed by the FJ model. We propose to construct an SFG that relates the influence of the stubborn agents to the network interactions using the steady-state behaviours. Thereafter, the SFG is reduced to a directed acyclic graph using the index residue reduction. Such a reduction allows for an easy determination of the change in the influence of the agents. For a class of strongly connected networks, this procedure allows us to re-structure the SFG such that all the nodes are distributed into level sets. Depending on the locations of the stubborn agents, the level sets are further classified into three regions as shown in Fig. \ref{fig:Lvel_distribution}. Through rigorous analysis, we show that for edge modifications carried out in region $1$, no change occurs in the influence of any of the stubborn agents. Hence, we know \textit{where not to add an edge in the network}. We also present the suitable edge modifications in regions $1$ and $2$ which can increase/reduce the influence of a desired stubborn agent.

The proposed graphical solution is simple, yet can be extremely useful in minimising energy/expenses in practical applications. Further, the set of feasible edge modifications can be substantially reduced through this approach; making it more tractable than optimisation-based frameworks. However, the topological conditions for ensuring non-redundant edge modifications still need to be explored in regions $2$ and $3$. In future, we also plan to analyse the scenario of multiple stubborn agents competing for influence and also extend our results to a wider class of networks.



\section*{APPENDIX A}
\subsection{Index Residue Reduction of the SFG}
\label{Sec:Index_Residue_Red}
The SFG $\mathcal{G}_s$ of a network satisfying Assumption \ref{Assump:index} has only one index node. Therefore, we apply the index-residue reduction procedure on $\mathcal{G}_s$ to determine its gain. 
First, we replace each self-loop in $\mathcal{G}_s$ with an equivalent branch. This results in an intermediary graph $\mathcal{\Tilde{G}}_s$, which has the same nodes and edges as $\mathcal{G}_s$. However, the branch gain of any edge $(i,j)$ in $\mathcal{\Tilde{G}}_s$ becomes $g_{j,i}/(1-g_{i,i})$
if $i$ had a self-loop in $\mathcal{G}_s$. 
Now, we further reduce the network $\mathcal{\Tilde{G}}_s$ to $\mathcal{G}_s^1$ by index residue reduction. The network $\mathcal{G}_s^1$ contains only the residual nodes that include the index node $m$, the sources $S_1$ and $S_2$ and the sink $O$ in $\mathcal{G}_s$. For a branch $(\alpha,\gamma)$ in $\mathcal{G}_s^1$, the branch gain $g_{\gamma,\alpha}^1$ is the summation of path gains of residual paths from $\alpha \to \gamma$ in $\mathcal{\tilde{G}}_s$. The self-loop at $m$ in $\mathcal{G}_s^1$ is replaced with a branch of gain $1/(1-g_{m,m}^1)$.
\begin{figure}[h]
    \centering

\includegraphics[width=0.8\linewidth,height=1.75cm,keepaspectratio]{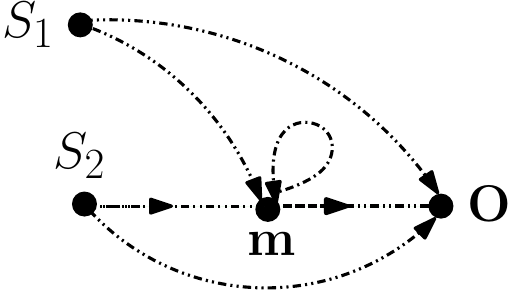}
       \caption{The reduced SFG $\mathcal{G}_s^1$. }
         \label{fig:gs2_variations_sinks}
\end{figure}
The gain $G_{O,S_i}$ of the SFG for source $S_i$ and sink $O$ is the sum of the path gains of the forward paths from $S_i$ to $O$ in $\mathcal{G}_s^1$, given by,
\begin{align}
\label{eqn:gain_eqn}
    G_{O,S_i}=\frac{g_{m,S_i}^1g_{O,m}^1}{g_{m,m}^1}+g_{O,S_i}^1
\end{align}
where $G_{O,S_i}$ is the influence centrality $c_{j}$ of stubborn agent $j$ when $S_i$ is associated with  $x_j(0)$.
\section*{Appendix B}
\label{Appendix:Thm_1}
\textbf{Proof of Theorem 1:} Since the edge modification $(a,b.d)$ in $\mathcal{G}$ is equivalent to the edge modification $(a,b,d)$ in $\mathcal{G}_s$, henceforth, we examine the consequences of altering the network in $\mathcal{G}_s$. 
We use the index residue reduction of the SFG $\mathcal{G}_s$, presented in Appendix-A, to determine the change in the influence of stubborn agents. 
It follows from eqn. \eqref{eqn:gain_eqn} that the influence centrality $c_{j}$ changes only if the branch gain of a branch in $\mathcal{G}_s^1$ changes due to $(a,b,d)$. In this proof, we will show that the gains of none of the branches change.

Recall that the branch gain of a branch $(e,f)$ in $\mathcal{G}_s^1$ is the summation of path gains of residual paths from residual node $e$ to residual node $f$ in $\mathcal{\tilde{G}}_s$. 
Since $a$ and $d$ do not have a direct path from any source,
a residual path that originates from $S_1$ or $S_2$ does not contain the branches $(a,b)$ or $(d,b)$.
Thus, the branch gain of branches $(S_j,m)$ and $(S_j,O)$ 
in $\mathcal{G}_s^1$ (see in Fig. \ref{fig:gs2_variations_sinks}) is unaffected for $j \in \{1,2\}$.

\begin{figure}[ht]
    \centering
    \begin{subfigure}{0.45\textwidth}
       \centering \includegraphics[width=0.7\textwidth,keepaspectratio]{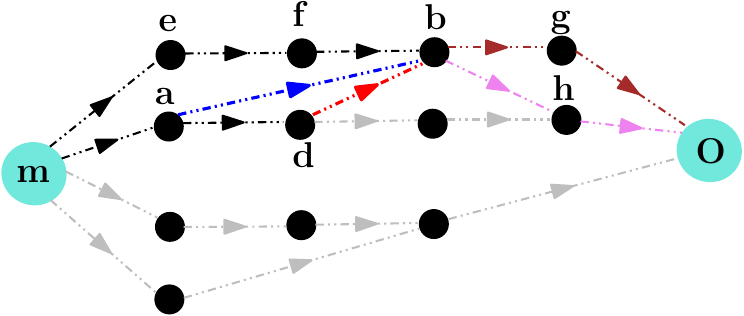}
        \label{fig:first_op}
    \end{subfigure}\hspace{0.2cm}
    \caption{Residual paths from $m$ to $O$ are shown. } 
    \label{Fig:Path_options}
\end{figure}

By definition, the in-degree of node $b$ in $\mathcal{G}$ and $\mathcal{G}_s$ remains unchanged after an edge modification. 
Thus, $g_{b,d}-\Tilde{g}_{b,d}=\Tilde{g}_{b,a}$ where $\Tilde{g}_{z,y}$ denotes the branch gain of $(y,z)$ in $\mathcal{G}_s$ after edge modification where $y,z \in V_s$.
The change in the branch gain of branch $(m,O)$ in $\mathcal{G}_s^1$ 
is denoted as $\Delta g_{O,m}^1=g_{O,m}^1-\tilde{g}_{O,m}^1$, where $\tilde{g}_{O,m}^1$ is branch gain of $(m,O)$ in $\mathcal{G}_s^1$ after edge modification. Then, $\Delta g_{O,m}^1$ is equal to the difference between the change in the path gains of the residual paths from $m \to O$ that pass via branch $(d,b)$ and the sum of path gains of new residual paths formed from $m \to O$ due to addition of $(a,b)$. Thus,
\begin{align}
\label{eqn:delta_g_2}
  &\Delta g_{O,m}^1= \sum_{P_{O,m}^{(d,b)}} \frac{g_{n_1,m}g_{n_2,n_1}... (g_{b,d}-\Tilde{g}_{b,d})...g_{O,n_r}}{(1-g_{m,m})(1-g_{n_1,n_1})...(1-g_{n_r,n_r})} \nonumber \\ 
 &-\sum_{P_{O,m}^{(a,b)}} \frac{g_{n_1,m}g_{n_2,n_1}...\Tilde{g}_{b,a}...g_{O,n_r}}{(1-g_{m,m})(1-g_{n_1,n_1})...(1-g_{n_r,n_r})}  
\end{align}
where $P_{O,m}^{(a,b)}$ and $P_{O,m}^{(d,b)}$ is the set of residual paths from $m \to O$ passing via the branches $(a,b)$ and $(d,b)$, respectively, and $n_1,n_2,...,n_r$ are the nodes in $\mathcal{G}_s$. It follows from eqn. \eqref{eqn:delta_g_2} that every residual path that affects $\Delta g_{O,m}^1$ passes through node $b$. 
Fig. \ref{Fig:Path_options} shows a portion of a hypothetical SFG that contains residual paths from index nodes $m$ to $O$. 
In Fig. \ref{Fig:Path_options}, three direct paths from $b$ \textit{i.e.} $(m \to a \to b)$, $(m\to a \to d \to b)$,$(m\to e \to f \to b)$ are highlighted. Note that irrespective of the path followed from $m \to b$, the direct paths from $b \to O$ remain the same \textit{i.e} $(b \to g \to O)$ and $(b \to h \to O)$. Therefore, it can be seen that $\Delta g_{\mathcal{O}_i,m}^1$ depends only on the direct paths from $m \to b$. 
Since the direct paths from $b \to O$ remain identical irrespective of the path followed from $m \to b$, the sum of path gains of all direct paths from $b \to O$ is denoted by $K$. It is elementary to derive (and hence omitted) that
 eqn. \eqref{eqn:delta_g_2} can be written as:
\begin{align*}
  \Delta g_{z,m}^1= & K\bigg(-\sum_{P_{z,m}^{(a,b)}} \frac{g_{n_1,m}g_{n_2,n_1}...\Tilde{g}_{b,a}}{(1-g_{m,m})(1-g_{n_1,n_1})...(1-g_{a,a})}  -  \nonumber \\ 
 &\sum_{P_{z,m}^{(d,b)}} \frac{g_{n_3,m}g_{n_4,n_3}... (g_{b,d}-\Tilde{g}_{b,d})}{(1-g_{m,m})(1-g_{n_3,n_3})...(1-g_{d,d})} \bigg)  
\end{align*}

Using the property of level sets $\mathcal{L}_{1},...,\mathcal{L}_q$ defined in Lemma \ref{lemma:disjoint_levels}, we further simplify eqn. \eqref{eqn:delta_g_2}. 
If $s1>1$, then a node $j$ in set $\mathcal{L}_1$ has only index node $m$ as its in-neighbour  and its branch gain $g_{j,m}=1$ if $j$ has no-self loops in $\mathcal{G}_s$. In case $j$ has self loops, $g_{j,m}<1$ but ${g_{j,m}}/({1-g_{j,j}})=1$. 
Similarly, if $s1 > 2$, then a node $j$ in $\mathcal{L}_2$ has only nodes in $\mathcal{L}_1$ and $m$ as its in-neighbours. Thus, the sum of path-gains of direct paths from $m \to j$ in $\mathcal{\tilde{G}}_s$ is,
\begin{align*}
  \sum_{k \in N_j^{in} \setminus \{m\}} \frac{g_{j,k}g_{k,m}}{(1-g_{k,k})(1-g_{m,m})}+\frac{g_{j,m}}{1-g_{m,m}} =  \frac{1-g_{j,j}}{1-g_{m,m}}
\end{align*}
The second equality holds
because the nodes in $N_j^{in} \setminus \{m\}$ are from set $\mathcal{L}_1$, we know that $\frac{g_{k,m}}{(1-g_{k,k})}=1$ for $k \in \mathcal{L}_1$. 
Consider a case in $\mathcal{\tilde{G}}$ when the sum of path gains of direct paths from $m$ to a node $k$ in set $\mathcal{L}_l$ for $l \in \{1,...,h\}$ is equal to ${(1-g_{k,k})}/{(1-g_{m,m})}$ and $s1>h$. Next, we prove that this condition holds for a node in set $\mathcal{L}_{h+1}$ as well if $s1>h+1$. 
Let node $j \in \mathcal{L}_{h+1}$.  Then, 
\begin{align}
\label{eqn:sum_direct_paths}
 & \mathcal{S}_{j,m}=  \sum_{k \in N_{j}^{in} \setminus \{m\}}\frac{g_{j,k}}{1-g_{k,k}} \mathcal{S}_{k,m} +\frac{g_{j,m}}{1-g_{m,m}} \nonumber \\ 
 &= \frac{1}{(1-g_{m,m})} \bigg( \sum_{k \in N_{j}^{in} \setminus \{m\}}g_{j,k} +g_{j,m}\bigg)=\frac{1-g_{j,j}}{1-g_{m,m}}
\end{align}
where $\mathcal{S}_{u,m}$ denotes the sum of the path gains of the direct paths from node $m$ to any node $u \in V_s$.
The second equality holds because the in-neighbours of node $j$ in $\mathcal{L}_{h+1}$ include nodes from sets $\mathcal{L}_1,...,\mathcal{L}_h$ and $m$, and we know that $\mathcal{S}_{k,m}=\frac{1-g_{k,k}}{1-g_{m,m}}$ for $k \in N_{j}^{in} \setminus \{m\}$. 
Therefore, by mathematical induction, it follows that for a node $j$ in $\mathcal{L}_{h+1}$ eqn. \eqref{eqn:sum_direct_paths} holds, if it holds for any node in
set $\mathcal{L}_{l}$ for $l=\{1,...,h\}$ where $h \in \mathbb{N}$ and $s1>h+1$. Eqn. \eqref{eqn:sum_direct_paths} also holds for a node $j \in \mathcal{L}_h$ where $h \geq s1$, if it holds for all of the in-neighbours of $j$. For Example, nodes in Fig. \ref{fig:Lvel_distribution} with red crosses.
Therefore, it follows that the relation in eqn. \eqref{eqn:sum_direct_paths} holds for a node $j$ only when no direct path from any source to $j$ exists. By the conditions in Theorem \ref{cor_1}, the nodes $a$ and $d$ in $(a,b,d)$
satisfy eqn. \eqref{eqn:sum_direct_paths}.
As a result, we can write eqn. \eqref{eqn:delta_g_2} as,
\begin{align*}
 & \Delta g_{O,m}^1=K\bigg(-\frac{\Tilde{g}_{b,a}}{1-g_{a,a}} \mathcal{S}_{a,m}+\frac{(g_{b,d}-\Tilde{g}_{b,d})}{(1-g_{d,d})}\mathcal{S}_{d,m}\bigg)  \\
  &=K\bigg(-\frac{\Tilde{g}_{b,a}}{1-g_{a,a}} \frac{1-g_{a,a}}{1-g_{m,m}} +\frac{(g_{b,d}-\Tilde{g}_{b,d})}{(1-g_{d,d})}\frac{1-g_{d,d}}{1-g_{m,m}}\bigg)=0 
\end{align*}
Thus, no change occurs in branch gain $g_{O,m}^1$. Following the same procedure, it can be proved that the branch gain $g_{m,m}^1$ is also unchanged. Since the branch gains of none of the branches in $\mathcal{G}_s^1$ are altered, it follows from eqn. \eqref{eqn:gain_eqn} that the influence of both the stubborn agents remains unchanged. \QED

\textbf{Proof of Corollary \ref{Cor_two}.}
As in the proof of Theorem \ref{cor_1}, we determine the change in the influence centrality $c_{j}$ of stubborn agent $j$ by 
calculating the changes in the gains of the branches in $\mathcal{G}_s^1$ resulting from edge modifications in $\mathcal{G}_s$. 

Given that the applied edge modification $(a,b,d)$ is such that $a$ does not have a direct path from any source in $\mathcal{G}_s$ and $d$ has a direct path  only from  $S_1$.
Under the given conditions, neither $(a,b)$ nor $(d,b)$ lies on any residual paths from $S_2$. Thus, the gain of branches in $\mathcal{G}_s^1$ from $S_2$ remains unchanged. On the other hand,
the gains of branches in $\mathcal{G}_s^1$ from $S_1$ and $m$ can get altered because $S_1$ has a direct path to $d$  and $m$ has a direct to every non-residual path in $\mathcal{G}_s$ due to Assumption \ref{Assump:index}. Next, we consider the effect on branches from $S_1$ due to the edge modification $(a,b,d)$.

A direct path exists from $S_1 \to d$, thus, branch $(d,b)$ lies on at least one residual path from $S_1 \to m$ in $\mathcal{G}_s$ but $(a,b)$ does not. Thus, it can be determined that the branch gain $g_{m,S_1}^1$ decreases. Similarly, it follows that 
the branch gain $g_{O,S_1}^1$ of $(S_1,O)$ also decreases. Node $m$ has a direct path to all the non-residual nodes in $\mathcal{G}_s$. Thus, the branches from $m$ in $\mathcal{G}_s^1$ are affected by both by addition of $(a,b)$ and the reduction of branch gain of $(d,b)$ in $\mathcal{G}_s$.
Given that node $a$ does not have a direct path from either of the sources, it follows from the proof of Theorem \ref{cor_1} 
that $\mathcal{S}_{a,m}=(1-g_{a,a})/(1-g_{m,m})$. However, since $d$ has a direct path from  $S_1$, so  $\mathcal{S}_{d,m}<(1-g_{d,d})/(1-g_{m,m})$. 
Thus, the change of branch gain of branch $(m,O)$ in $\mathcal{G}_s^1$ is
$\Delta g_{O,m}^1 =\frac{K}{{1-g_{m,m}}}(-\Tilde{g}_{b,a} +\rho ({g}_{b,d}- \Tilde{g}_{b,d})) 
    = \frac{K\Tilde{g}_{b,a}}{{1-g_{m,m}}}(-1+\rho)<0$, where $\rho<1$. 
So, the branch gain of $(m,O)$ increases. A similar analysis shows that the loop gain of self-loop $g_{m,m}^1$ also increases. 

\begin{figure}[h]
        \centering
            \includegraphics[width=0.4\linewidth]{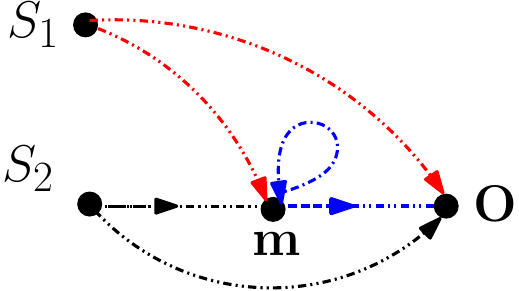}
   \caption{Effect of edge modifications on branch gains of $\mathcal{G}_s^1$. }
  \label{fig:proof_fig}
\end{figure}

The alterations in branches of $\mathcal{G}_s^1$ due to $(a,b,d)$ are shown in Fig. \ref{fig:proof_fig}. The branches
in red, blue and black denote a decrease, an increase and no change in the branch gains, respectively. 
It is simple to observe from eqn. \eqref{eqn:gain_eqn} that the gain of SFG with source $S_2$ and sink $O$ increases due to edge modification. Thus, the influence centrality of the stubborn agent associated with source $S_2$ increases. Since $\mathbf{c}^T\mathbb{1}_n=1$, the influence centrality of the stubborn agent associated with source $S_1$ decreases. \QED
  
\bibliographystyle{IEEEtran}
\bibliography{references}
\end{document}